\documentclass[%
reprint,
 amsmath,amssymb,
 longbibliography,
 aps,
]{revtex4-1}

\usepackage[english]{babel}
\usepackage[normalem]{ulem}
\usepackage{graphicx}
\usepackage{dcolumn}
\usepackage{bm}
\usepackage{xcolor}
\usepackage{tensor}
\usepackage{comment}
\usepackage{verbatim}

\newcommand{\eps}{\varepsilon}

\begin{document}


\title{Emergence of zero modes in disordered solids under periodic tiling}

\author{R. Cameron Dennis$^*$, Varda F. Hagh$^{*,\dagger}$, Eric I. Corwin$^*$}
\affiliation{$^*$Department of Physics and Materials Science Institute, University of Oregon, Eugene, Oregon 97403, USA \\ $^\dagger$James Franck Institute, University of Chicago, Chicago, IL 60637, USA}

\date{\today}

\begin{abstract}
In computational models of particle packings with periodic boundary conditions, it is assumed that the packing is attached to exact copies of itself in all possible directions. The periodicity of the boundary then requires that all of the particles' images move together. An infinitely repeated structure, on the other hand, does not necessarily have this constraint. As a consequence, a jammed packing (or a rigid elastic network) under periodic boundary conditions may have a corresponding infinitely repeated lattice representation that is not rigid or indeed may not even be at a local energy minimum. In this manuscript, we prove this claim and discuss ways in which periodic boundary conditions succeed to capture the physics of repeated structures and where they fall short.
\end{abstract}

\maketitle
\section*{Introduction}

Periodic boundary conditions are a mainstay in the theoretical and computational study of condensed matter systems as they ameliorate or eliminate many finite-size effects, allowing one to infer bulk behavior from small systems. In simulations of physical systems ``periodic boundaries'' treat the system under study as a unit cell interacting with exact copies of itself in every direction.
It is easy to take periodic boundary conditions for granted and think of them as capturing all of the physics of a repeated tiling with infinite \textit{independent} copies of the original system. But this misses a crucial distinction: a system with periodic boundary conditions requires that all of the particle images in a repeated tiling move in concert, whereas infinitely repeated structures have no such constraint. This distinction between free and periodic boundary conditions is important because it has a profound impact on one of the most fundamental properties of networks and packings: rigidity.

In this paper, we provide a set of mathematical arguments and proofs, as well as numerical results, that demonstrate how the rigidity criterion for first-order rigid systems~\cite{damavandi_energetic_2022} with periodic boundaries changes when the system is repeated in space. In particular, in a marginally rigid system (i.e.~a system near the rigidity transition point), duplicating the system and considering two attached copies as the new unit cell, introduces new floppy modes that can break the rigidity of the entire system. This implies that by tiling space with such systems, one cannot produce an infinitely large system that retains its rigidity. 

For a finite-sized packing (or network), it is natural to ask how many constraining contacts (or bonds) are needed for rigidity. The Maxwell-Calladine rule provides an answer~\cite{f.r.s_l._1864, calladine_buckminster_1978, malestein_generic_2013}:
\begin{equation}
F - S = Nd - N_c,
\label{eq: maxwell_equation}
\end{equation}
where $F$ is the total number of floppy modes (or zero modes), $N$ is the number of particles (or nodes), $d$ is the spatial dimension (making $Nd$ the total number of degrees of freedom), $N_c$ is the number of constraining contacts (or bonds), and $S$ is the number of redundant constraints which is equal to the number of states of self-stress. States of self-stress include all the possible ways a system can support contact forces while being at mechanical equilibrium. For a physical system to be rigid~\cite{calladine_first-order_1991, jacobs_generic_1995, jacobs_algorithm_1997, damavandi_energetic_2022}, it must only have trivial rigid motions as floppy modes. For instance, a $d$ dimensional finite system is considered rigid if it only has $F = d + d (d-1)/2$ floppy modes including $d$ translations and $d(d-1)/2$ rotations. For infinitely large systems, the only trivial rigid motions are the translations. Thereby, a $d$ dimensional packing (or network) that is under periodic boundary conditions or is infinitely repeated in space, can only have $d$ floppy modes when rigid.

Note that the Maxwell-Calladine constraint counting rule is not a suitable proxy for measuring rigidity in all types of physical systems. For instance, in second-order rigid systems, such as under-constrained networks that rigidify under tension, Eq.~(\ref{eq: maxwell_equation}) cannot be used to describe the rigidity~\cite{connelly_second-order_1996,alexander_amorphous_1998, holmes-cerfon_almost-rigidity_2021, damavandi_energetic_2022, damavandi_energetic_2022-1}. Another example where this constraint counting method fails is in systems with shear degrees of freedom or special symmetries (such as square or Kagome lattices), where the alignment of states of self-stress can lead to internal floppy modes that are \textit{not} included in the Maxwell-Calladine count~\cite{guest_determinacy_2003, owen_frameworks_2010}. However, for all the systems studied in this paper, including jammed packings of soft particles and elastic networks, Eq.~(\ref{eq: maxwell_equation}) is a sufficient proxy for measuring rigidity. 

\begin{figure}[]
\includegraphics[width=0.475\textwidth]{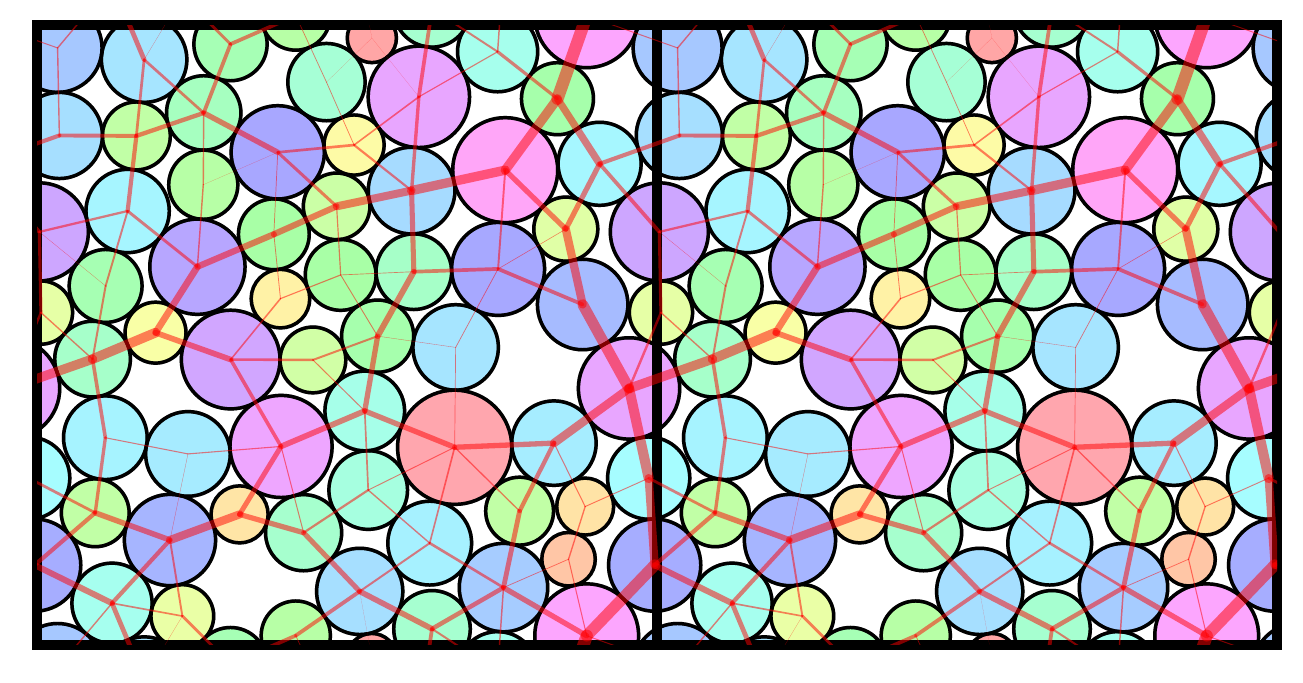}
\caption{A packing of spheres that has been duplicated. The red lines display a state of self-stress. The thickness of each line represents the magnitude of the stress on the corresponding bond. Replicating the state of self-stress for the original system gives a state of self-stress for the duplicated system.}\label{fig:SS}
\end{figure}

Suppose we have a first-order rigid system with periodic boundaries, $d$ floppy modes, and $S$ states of self-stress. Note that first-order rigidity in systems with periodic boundary conditions implies that the number of constraints is greater than or equal to the number of degrees of freedom. By duplicating the unit cell and attaching the copy to the cell across one of the boundaries (as demonstraited in \ref{fig:SS}), both the number of particles and the number of constraints double in size. How does this impact the number of floppy modes and states of self-stress? One might naively assume that both $F$ and $S$ should also double to satisfy Eq.~(\ref{eq: maxwell_equation}). However, this would imply that the new system has $2d$ floppy modes and thus is non-rigid. Does this mean that rigid systems do not remain rigid under duplication? This is not a generally true statement, because many crystalline structures (such as triangular lattices) are rigid under infinite tiling. In the following sections, we show that changes in the number of floppy modes and states of self-stress when a periodic rigid structure undergoes duplication (or more generally tiling) are non-trivial and depend on multiple factors, including the initial number of states of self-stress in the unit cell.


To show the impact of repeating a physical system on its rigidity and how it compares to the rigidity under periodic boundaries, we first examine the mathematical structure of periodic boundary conditions in jammed packings of soft athermal particles and their underlying spring networks. Note that in jammed packings of soft particles, there are almost always prestress forces present in the system, while networks can be either stressed or unstressed. In addition, states of self-stress in such systems are system spanning, meaning that they involve all of the contact bonds.
In the absence of any prestress, the rigidity of a system can be fully captured by its rigidity matrix, $\mathbf{R}$, which is the matrix of first derivatives of constraints $h$ (overlaps in soft particle packings and bond lengths in spring networks) with respect to degrees of freedom, $R_{\alpha, i} = \frac{\partial h_{\alpha}}{\partial x_{i}}$. $\mathbf{R}$ is a $N_c\times Nd$ dimensional matrix (with $N_c$ being the number of constraints), where each row, $\alpha$, represents a pair of interacting particles (or connected nodes) and every $d$ consecutive columns correspond to a particle. The rigidity matrix relates changes in degrees of freedom, $\delta \mathbf{X}$, to changes in the constraints, $\mathbf{\Delta}$, via $\mathbf{R} \delta \mathbf{X} = \mathbf{\Delta}$. It is trivial to show that
the right null space of $\mathbf{R}$ represents floppy modes in the system that correspond to motions that do not change the constraints~\cite{f.r.s_l._1864, calladine_buckminster_1978, malestein_generic_2013}. 
In the presence of prestress forces, however, one must compute the null space of the Hessian matrix (that includes a negative definite prestress term) to find the floppy modes~\cite{f.r.s_l._1864, calladine_buckminster_1978, malestein_generic_2013}. The Hessian, $\mathbf{H}$, is the second derivative of energy function with respect to all degrees of freedom, $\mathbf{H}_{ij}^{\alpha \beta} = \frac{\partial^2 U}{\partial x_i^{\alpha} \partial x_j^{\beta}}$. When the prestress forces are small \textit{and} the system is mechanically stable, the nullity of the rigidity matrix can still show the number of floppy modes in the system.

Following the more statistical findings of Goodrich et al.~\cite{goodrich_stability_2013, schoenholz_stability_2013}, who show that infinitely tiled two-dimensional disk packings can lead to anomalously low-frequency modes, we demonstrate that infinitely tiled jammed packings of soft spheres in any dimension can not only have anomalously low-frequency modes but also new zero or even negative modes when they are sufficiently close to critical jamming, i.e. with fewer than $d$ states of self-stress.
With this goal in mind, we prove the following theorems and arguments (presented here in simplified terms):

\begin{enumerate}
\item[TI:] An unstressed jammed sphere packing with periodic boundary conditions and fewer than $d$ states of self-stress will not remain jammed after tiling space,
\item[TII:] When there are $S < d$ states of self-stress in an unstressed jammed packing or spring network, duplicating the system across any boundary will introduce at least $d - S$ new zero modes to the system,
\item[AI:] Tilings of amorphous over-jammed packings of soft spheres (with prestresses), even with $d$ or more states of self-stress, typically have unjamming motions,
\item[AII:] Unstressed networks with $d$ or more states of self-stress typically \textit{do} have a rigid infinite lattice representation, but we show that there exist special counter-examples which do not,
\item[TIII:] The bulk elastic properties of an infinitely repeated packing are fully captured by periodic boundary conditions.
\end{enumerate}

\section*{Theorem I:}

\textbf{For unstressed jammed packings (or unstressed spring networks) with $S<d$ states of self-stress, the corresponding packing (or network) that is duplicated across boundary $x$ will not be rigid under the assumption that both the original and duplicated systems have periodic boundary conditions}.




\subsection*{Proof}

We proceed with a proof by contradiction. Imagine a jammed packing with no prestresses. $F=d$ trivial floppy modes, and $S<d$ states of self-stress. Assume that the duplicated system is jammed and therefore has $F'=d$ trivial floppy modes and $S'$ states of self-stress. These trivial floppy modes are simply the modes that correspond to translating each particle by the same amount in the same direction. From the Maxwell-Calladine constraint counting rule in Eq.~\ref{eq: maxwell_equation},
\begin{align}
Nd-N_c&=F-S\\
\textrm{and~}2Nd-2N_c&=F'-S'
\end{align}
where $Nd$ is the total number of degrees of freedom and $N_c$ is the number of contacts in the original packing. Duplicating a packing across one of the boundaries will indeed double both the number of particles and the number of contacts. We can therefore state that $F'-S'=2(F-S).$ If we substitute our assumption that $F=F'=d$ and solve for $S',$ then
\begin{align}
S'=2S-d.\label{eq:sp}
\end{align}

Now if we have only one state of self-stress in the original system, we can simply replicate it to find a state of self-stress in the duplicated system~\cite{borcea_liftings_2015} (as shown in Fig.~\ref{fig:SS}). If applying a set of stresses to the contact bonds of the original packing leads to force balance, then replicating those stresses to a duplicated packing must also lead to force balance. This means that we can automatically find $S$ states of self-stress for the duplicated system. Note that orthogonality of these $S$ states of self-stress is preserved in the doubling procedure. However, it is possible to find additional states of self-stress for the duplicated packing which are not captured by this trivial doubling. Therefore, $S'\geq S.$ Substituting this result into Eq.~\ref{eq:sp} gives:
\begin{align}
2S-d&\geq S \\
S&\geq d.
\end{align}

We assumed at the beginning of this proof that the original and duplicated packings are jammed and that $S<d.$ We have thus reached a contradiction. Any jammed packing or spring network with $S<d$ states of self-stress, when duplicated across any boundary, must display an unjamming motion. 

\section*{Theorem II:}
\textbf{For an unstressed jammed packing (or spring network) with $S<d$ states of self-stress, duplicating the packing (or network) across boundary $x$ produces at least $d - S$ emergent floppy modes in the system.}

\subsection*{Definitions}
To understand the following proof, we must first define the rigidity matrix for a spring network or packing. As discussed in the introduction, the rigidity matrix relates the forces applied to the particles with the resulting stresses on the contacts. Consider the displacement vector of particle $i$ to be given by $\vec{x}_i$ and the unstressed bond between particles $i$ and $j$ be given by the normalized vector, $\vec{n}_{ij}.$ From this, we can define the stress of the bond (i.e. the lengthening or shortening of said bond) to be given by the dot product
\begin{align}
\vec{n}_{ij}\cdot \left(\vec{x}_j-\vec{x}_i\right).
\end{align}

If we let the first index of our rigidity matrix be the bond between particles $i$ and $j$ and let the second index be the degree of freedom $k\gamma,$ then
\begin{align}
n_{ij}^{\gamma}\left(x_j^{\gamma}-x_i^{\gamma}\right)=R_{\langle ij\rangle\left(k\gamma\right)}x_{k}^{\gamma}
\end{align}
where $k$ denotes the particle index and $\gamma$ denotes the dimension index. For this relationship to be correct, we define
\begin{align}
R_{\langle ij\rangle\left(k\gamma\right)}\equiv \left(\delta_{jk}-\delta_{ik}\right)n_{ij}^{\gamma}
\end{align}
where $\delta$ is the Kronecker delta function.

\subsection*{Proof}
To prove this stronger theorem, we first need to understand how the singular values of the rigidity matrix in a duplicated system compare to the singular values of the rigidity matrix in the original system. To achieve this, we introduce the following lemma. 
\section*{Lemma I}
\textbf{The rigidity matrix of any unstressed jammed packing (or network) under periodic boundary conditions can be written as 
\begin{align*}
    R=\begin{pmatrix}
    R_{c} & R_{p} \\
    0 & R_{b2}+R_{b1}
    \end{pmatrix}
\end{align*}
where $\begin{pmatrix}
0 & R_{b2}+R_{b1}
\end{pmatrix}$ corresponds to the contacts crossing boundary $x.$ We denote the singular values of this matrix to be $\left\{\sigma_{i}\right\}.$ 
We introduce a new matrix called the {\it Doubled Hessian Compliment Matrix}, 
\begin{align*}
    R_{\text{DHC}}\equiv \begin{pmatrix}
    R_{c} & R_{p} \\
    0 & R_{b2}-R_{b1}
    \end{pmatrix}
\end{align*}
and denote the singular values of the matrix to be $\left\{\epsilon_{i}\right\}.$ Under these definitions, duplicating the packing across boundary $x$ results in a system where the rigidity matrix has singular values $\left\{\sigma_{i}, \epsilon_{i}\right\}$.
}
\subsection*{Proof}
Consider $b$ contact bonds that cross boundary $x$, with $p$ particles involved in making these boundary contacts. We separate the rigidity matrix, $R$, into columns that do not involve these $p$ boundary particles and columns that do. We further separate $R$ into rows that do not involve these $b$ contacts and rows that do. This gives
 \begin{align}
    R=\begin{pmatrix}
    R_{c} & R_{p} \\
    0 & R_{b2}+R_{b1}
    \end{pmatrix}
\end{align}
where $R_c$ (which has shape $(N_c-b)\times (Nd-p)$) represents the contacts that are formed between non-boundary particles, $R_p$ ($(N_c-b)\times p$) represents contacts that involve the $p$ boundary particles but do not cross the boundary themselves, $R_{b1}$ ($b\times p$) are the rows that involve boundary contacts where the rightmost vectors in each row are zero, and $R_{b2}$ ($b\times p)$ are the rows of the boundary contacts where the leftmost vectors in each row are zero. The reordering of columns of the rigidity matrix corresponds simply to reindexing the particles and the reordering of rows of the rigidity matrix corresponds to reindexing the bonds. This $0$ block comes from our definition of non-boundary particles having no boundary contacts.

Replicating the system across boundary $x$ results in a new system with rigidity matrix $R_{\text{D}}$, that can be written as
\begin{align}
    R_{\text{D}}=\begin{pmatrix}
    R_c & R_p & 0 & 0 \\
    0 & R_{b2} & 0 & R_{b1} \\
    0 & 0 & R_c & R_p \\
    0 & R_{b1} & 0 & R_{b2}
    \end{pmatrix}.
\end{align}

If we separate this matrix into four $2\times 2$ blocks, we notice that the top left and bottom right blocks simply describe the interaction of each replica with itself. When the configuration is duplicated across boundary $b_1,$ the bonds that cross this boundary in the original packing do not cross it in the duplicated packing. Therefore, these diagonal blocks are the same as $R$ except that we remove the $R_{b1}$ term. In contrast, the off-diagonal $2\times 2$ blocks of $R_{\text{D}}$ only contain the $R_{b1}$ term.
 
If we consider $\vec{u}=\begin{pmatrix}
\vec{u}_{1} \\ \vec{u}_{2}
\end{pmatrix}$ and $\vec{v}=\begin{pmatrix}
\vec{v}_{1} \\ \vec{v}_{2}
\end{pmatrix}$ to be left and right singular vectors (respectively) for $R$ with corresponding singular value $\sigma$ such that $\vec{u}^TR\vec{v}=\sigma,$ then we can demonstrate that $\frac{1}{\sqrt{2}}\begin{pmatrix}\vec{u} \\ \vec{u}\end{pmatrix}=\frac{1}{\sqrt{2}}\begin{pmatrix}
\vec{u}_{1} \\ \vec{u}_{2} \\ \vec{u}_{1} \\ \vec{u}_{2}
\end{pmatrix}$ and $\frac{1}{\sqrt{2}}\begin{pmatrix}\vec{v} \\ \vec{v}\end{pmatrix}=\frac{1}{\sqrt{2}}\begin{pmatrix}
\vec{v}_{1} \\ \vec{v}_{2} \\ \vec{v}_{1} \\ \vec{v}_{2}
\end{pmatrix}$ are left and right singular vectors for $R_{\text{D}}$ with singular value $\sigma.$ Note that these vectors maintain their orthonormality condition. Shown explicitly,
\begin{align*}
\frac{1}{2}\begin{pmatrix}
\vec{u}^T & \vec{u}^T\end{pmatrix}&R_{\text{D}}\begin{pmatrix}
\vec{v} \\ \vec{v}
\end{pmatrix}\\
&=\frac{1}{2}\begin{pmatrix}
\vec{u}^T & \vec{u}^T\end{pmatrix}\begin{pmatrix}
    R_c & R_p & 0 & 0 \\
    0 & R_{b2} & 0 & R_{b1} \\
    0 & 0 & R_c & R_p \\
    0 & R_{b1} & 0 & R_{b2}
    \end{pmatrix}\begin{pmatrix}
\vec{v}_{1} \\ \vec{v}_{2} \\ \vec{v}_{1} \\ \vec{v}_{2}
\end{pmatrix}\\
&=\frac{1}{2}\begin{pmatrix}
\vec{u}^T & \vec{u}^T\end{pmatrix}\begin{pmatrix}
\begin{pmatrix}R_{c} & R_{p}\\
0 & R_{b2}+R_{b1}\end{pmatrix}\begin{pmatrix}
\vec{v}_1\\
\vec{v}_2 
\end{pmatrix}\\
\begin{pmatrix}R_{c} & R_{p}\\
0 & R_{b2}+R_{b1}\end{pmatrix}\begin{pmatrix}
\vec{v}_1\\
\vec{v}_2 
\end{pmatrix}
\end{pmatrix}\\
&=\frac{1}{2}\begin{pmatrix}
\vec{u}^T & \vec{u}^T\end{pmatrix}\begin{pmatrix}
R\vec{v}\\
R\vec{v}
\end{pmatrix}\\
&=\sigma.
\end{align*}
This means $\left\{\sigma_i\right\}$ are also singular values for $R_{\text{D}}.$ This is not surprising because a particle and its replica have the same displacement vectors in the eigenvectors that correspond to these singular values. Therefore, these are the modes that correspond to the particles moving in concert with their replicas. In a similar fashion, consider $\vec{x}$ and $\vec{w}$ to be the left and right singular vectors for $R_{\text{DHC}}$ with singular value $\epsilon.$ We can now show that $\frac{1}{\sqrt{2}}\begin{pmatrix}
\vec{x} \\ -\vec{x}
\end{pmatrix}$ and $\frac{1}{\sqrt{2}}\begin{pmatrix}
\vec{w} \\ -\vec{w}
\end{pmatrix}$ are left and right singular vectors for $R_\text{D}$ with singular value $\epsilon.$ Again, note that these vectors are orthonormal and consider,
\begin{align*}
\frac{1}{2}\begin{pmatrix}
\vec{x}^T & -\vec{x}^T\end{pmatrix}&R_{\text{D}}\begin{pmatrix}
\vec{w} \\ -\vec{w}
\end{pmatrix}\\
&=\frac{1}{2}\begin{pmatrix}
\vec{x}^T & -\vec{x}^T\end{pmatrix}\begin{pmatrix}
R_\text{DHC}\vec{w}\\
-R_\text{DHC}\vec{w}
\end{pmatrix}\\
&= \vec{x}^TR_\text{DHC}\vec{w}\\
&=\epsilon.
\end{align*}
This means that $\left\{\epsilon_i\right\}$ are also singular values for $R_{\text{D}}.$ To complete the proof, notice that 
\begin{align*}
\frac{1}{2}\begin{pmatrix}\vec{u}^T & \vec{u}^T\end{pmatrix}\begin{pmatrix}\vec{x}^T \\ -\vec{x}^T\end{pmatrix}=0
\end{align*}
and
\begin{align*}
\frac{1}{2}\begin{pmatrix}\vec{v}^T & \vec{v}^T\end{pmatrix}\begin{pmatrix}\vec{w}^T \\ -\vec{w}^T\end{pmatrix}=0
\end{align*}
Since $R_{\text{D}}$ has precisely twice as many singular values as $R$ and since the above orthogonality condition holds, all of the singular values for $R_{\text{D}}$ must be given by $\left\{\sigma_{i},\epsilon_{i}\right\}.$ 

Now we can use this information to prove the theorem. From the rank-nullity theorem, we know that for a system with $d$ trivial floppy motions,

\begin{align*}
\text{rank}(R) = Nd -d.
\end{align*}
On the other hand, the maximum rank that the $R_{\text{DHC}}$ matrix can have is the minimum of its number of rows and columns, $\text{max(rank}(R_\text{DHC})) = \text{min}(Nd, N_c)$. We later show how the rank of the $R_\text{DHC}$ can be computed in Lemma II. Replacing $N_c$ with $Nd - d + S$ gives
\begin{align*}
\text{max(rank}(R_\text{DHC})) = \text{min}(Nd, Nd -d + S).
\end{align*}
For a system with $S < d$ states of self-stress, $\text{max(rank}(R_\text{DHC}))$ is $Nd -d + S$. Thus $\text{rank}(R_\text{DHC}) \leq Nd -d + S$. This means,
\begin{align*}
\text{rank}(R)+\text{rank}(R_\text{DHC})  \leq 2Nd -2d + S.
\end{align*}
We know from Lemma I that $R$ and $R_\text{DHC}$ capture all of the behavior of $R_\text{D}$ in first-order rigid systems. Because the sets of right singular vectors for $R$ and $R_{\text{DHC}}$ make up all of the right singular vectors for $R_{\text{D}},$ we know that $\text{rank}(R_\text{D}) = \text{rank}(R) + \text{rank}(R_\text{DHC}).$ This gives
\begin{align*}
\text{rank}(R_\text{D}) \leq 2Nd -2d + S,
\end{align*}
which implies
\begin{align}
\label{eq:nullityOfR_D}
\text{nullity}(R_\text{D}) \geq 2d - S.
\end{align}
For the duplicated system to be fully rigid and jammed, $\text{nullity}(R_\text{D})$ must be equal to $d$. However, Eq.~\ref{eq:nullityOfR_D} shows that there are at least $d - S$ new zero modes appearing in the system. This proves the result that for a jammed packing or spring network with $S<d$ states of self-stress, duplicating the packing across boundary $x$ produces at least $d - S$ emergent floppy modes in the system.
\newline

So far, we have proven that unstressed amorphous systems with fewer than $d$ states of self-stress are not rigid upon replication. But what can we say about systems that are far from criticality and have more than $d$ states of self-stress? In the following, we first consider jammed packings of soft athermal spheres with prestress forces and show that the presence of prestress increases the probability of unjamming when the system is repeated infinitely. We then consider the case of unstressed systems and provide an argument for why these systems typically \textit{do} remain jammed under tiling.

\section*{Argument I:}
\textbf{Nearly all amorphous systems with fixed boundaries and non-zero prestress will destabilize under a sufficient number of replications.}
\subsection*{Definitions}
It can be shown that jammed packings of soft particles with more than $d$ states of self-stress, typically are marginally stable~\cite{hagh_transient_2022} and thereby represent saddle points on the energy landscape of the tiled system. We can show this by considering the Hessian in the momentum space, $H(\vec{q}),$ found through Bloch's theorem~\cite{kittel_introduction_2004}. For simplicity in calculations, we consider the interaction of a packing with its neighboring replicas. Each unit cell has $3^d-1$ neighboring cells in a tiling. We consider the interaction between a unit cell and its neighbor, $i,$ as $H_i.$ If we let the interaction between the unit cell and itself be $H_0,$ then the original Hessian, $H(\vec{0}),$ can be written as
\begin{align}
H(\vec{0})=H_0+\sum_{i=1}^{3^d-1}H_i.
\end{align}

\begin{figure}[]
\includegraphics[width=0.475\textwidth]{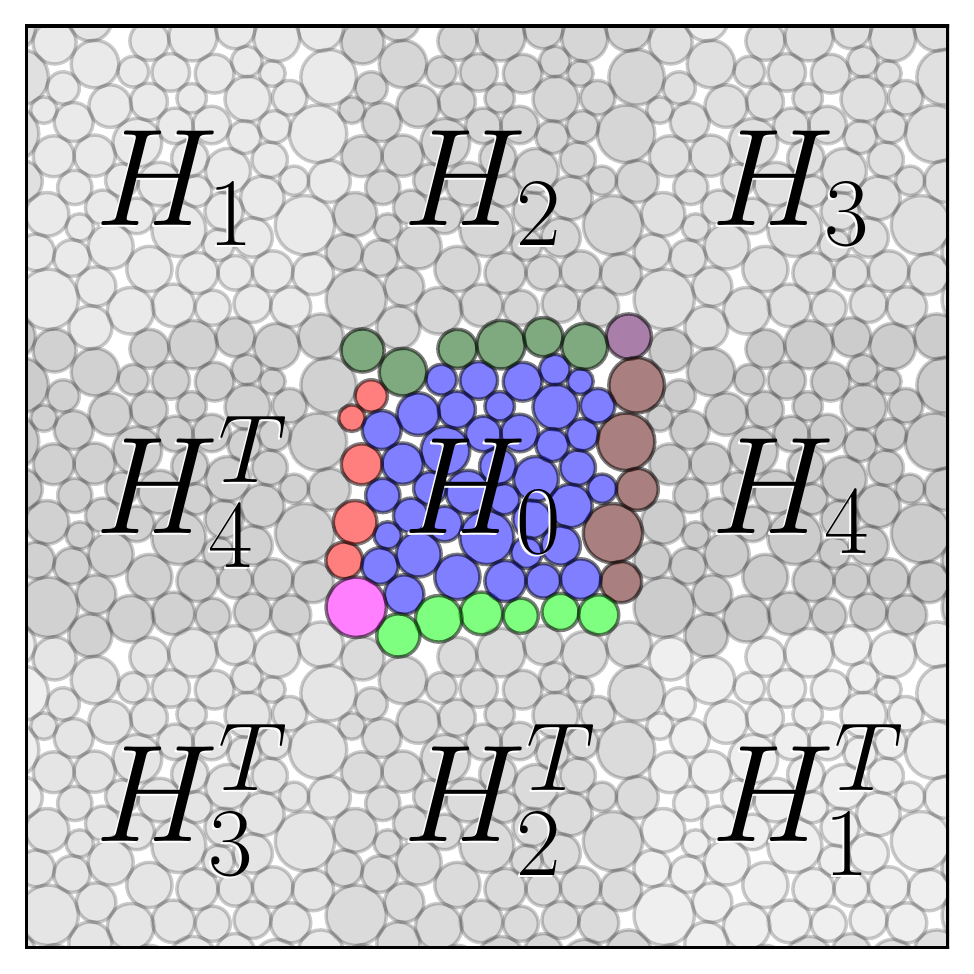}
\caption{The Hessian can be split into pieces corresponding to the interactions with neighboring copies of the unit cell. There are $(3^d+1)/2$ independent Hessian pieces which are shown in the center of their corresponding cells. Blue is used to represent those particles that do not interact with neighboring replicas, whereas the other colors represent interactions with neighboring replicas. The dark greens for example are those particles that interact with particles in the cell labeled $H_2.$ The magenta particle is the only particle that interacts with particles in the cell labeled $H_3^T.$ Notice that the orange and brown particles would be in contact for an unreplicated system. Therefore, we label the neighboring cells with $H_4^T$ and $H_4$ to account for this symmetry.}\label{fig:neighbors}
\end{figure}

This can be simplified further. If $H_j$ is the Hessian component with respect to the neighboring copy, $j$, and $H_k$ is the Hessian component with respect to the opposite neighboring copy, $k$, then $H_k=H_j^T$ as shown in Fig.~\ref{fig:neighbors}. This means that the Hessian only needs to be split into $(3^d+1)/2$ parts and
\begin{align}
H(\vec{0})=H_0+\sum_{i=1}^{(3^d-1)/2}\left(H_i+H_i^T\right).
\end{align}
In general, from Bloch's theorem, we can conclude that
\begin{align}
H(\vec{q})=H_0+\sum_{i=1}^{(3^d-1)/2}&\left[\left(H_i+H_i^T\right)\cos(\vec{q}\cdot \vec{r}_i)\right.\nonumber \\
&\left.\left(H_i-H_i^T\right)\sin(\vec{q}\cdot \vec{r}_i)\right]
\end{align}
where $\vec{r}_i$ is the $d$ dimensional vector corresponding to the position of cell $i$.

\begin{figure}[]
\includegraphics[width=0.475\textwidth]{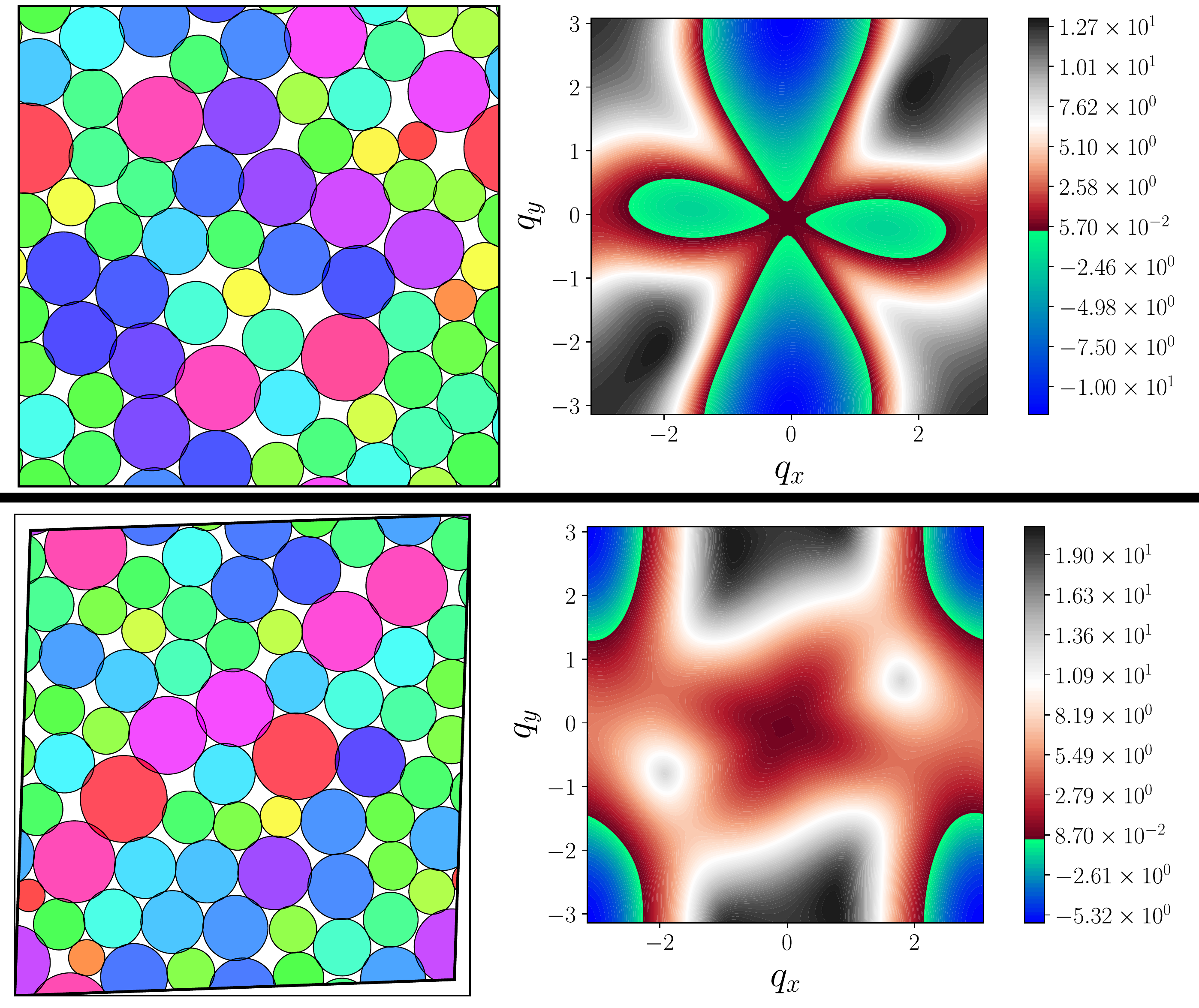}
\caption{\textbf{Top Left}: An over-jammed amorphous packing of $64$ harmonic soft spheres at packing fraction $\phi = 0.90$ with $30$ states of self-stress. \newline \textbf{Top Right}: The eigenvalues in the first branch of the momentum-space Hessian for this over-constrained packing. The blue and green colors represent the negative eigenvalues that correspond to perturbations that lower the energy.\newline \textbf{Bottom Left}: A shear stabilized packing of $64$ harmonic soft spheres at $\phi=0.90$ and $26$ states of self-stress.\newline \textbf{Bottom Right}: The corresponding contour plot of the first branch eigenvalues of the shear stabilized system's Hessian in momentum space.}\label{fig:branch0}
\end{figure}

\subsection*{Argument}
We can determine if a tiled packing remains jammed upon replication by looking at the eigenvalues of the first branch (lowest band of the momentum-space Hessian).  
For a packing that remains jammed when tiled, all of these eigenvalues should be greater than zero except for the trivial zero modes that come from $H(\vec{0}).$ Looking at a small jammed packing with $N = 64$ particles and $S = 30$ states of self-stress in 2D, we can see that the eigenvalues in the first branch are negative for certain values of momentum (see Fig.~\ref{fig:branch0}). A negative eigenvalue means there is a direction in which the particles can be perturbed that lowers the energy of the tiled system. This implies that the tiled packing, while in force balance, is not jammed.
Shear stabilized packings also may have negative modes in their first branch. An example is shown in the lower panel of Fig.~\ref{fig:branch0} with a shear stabilized packing of $N = 64$ soft harmonic particles and $S  = 26$ states of self-stress. These examples demonstrate that the tiling of an over-jammed packing might unjam when the number of tiles goes to infinity.

The above argument concerns jammed packings of soft spheres with non-zero prestresses. However, it turns out that in unstressed systems such as jammed packings of hard spheres or elastic networks with zero prestress, duplicating the system is unlikely to break its stability when there are more than $d$ states of self-stress. Remember that in such systems, the rigidity can be explored using the null space of the rigidity matrix.

\section*{Argument II:}
\textbf{For typical unstressed systems with $S\geq d$ states of self-stress, the corresponding system that is duplicated across boundary $x$ will typically be rigid and the system will typically remain rigid when tiled.}

\subsection*{Reasoning}
Let the singular value decomposition of $R$ be given by $R=\begin{pmatrix}
U_c \\ U_b
\end{pmatrix}^T\Sigma\begin{pmatrix}
V_c \\ V_p
\end{pmatrix}$ where $U_b$ represents the $b$ rows of the left unitary singular vector matrix which correspond to boundary bonds across boundary $x$ and $V_p$ represent the $p$ rows of the right unitary singular vector matrix which correspond to particles that do not have boundary bonds.
Let $X$ and $W$ be left and right unitary singular vector matrices for $R_\text{DHC}$ such that $R_{DHC}=X\Sigma_\text{DHC}W^T.$ Further, consider $\alpha$ and $\beta$ to be the change of the basis matrices such that
\begin{align*}
    W&=\begin{pmatrix}
V_c \\ V_p
\end{pmatrix}\alpha\\
    \text{and~} X&=\begin{pmatrix}
U_c \\ U_b
\end{pmatrix}\beta.
\end{align*}
Now consider
\begin{align*}
    \Sigma_\text{DHC}&=X^TR_\text{DHC}W\\
    &=\beta^T\begin{pmatrix}
U_c \\ U_b
\end{pmatrix}^TR_\text{DHC}\begin{pmatrix}
V_c \\ V_p
\end{pmatrix}\alpha\\
    &=\beta^T\begin{pmatrix}
U_c \\ U_b
\end{pmatrix}^T\left[R-\begin{pmatrix}
    0 & 0\\
    0 & 2R_{b1}
    \end{pmatrix}\right]\begin{pmatrix}
V_c \\ V_p
\end{pmatrix}\alpha\\
    &=\beta^T\left[\Sigma-\begin{pmatrix}
U_c \\ U_b
\end{pmatrix}^T\begin{pmatrix}
    0 \\ 2R_{b1}V_p
    \end{pmatrix}\right]\alpha\\
    &=\beta^T\left(\Sigma-2U_b^TR_{b1}V_p\right)\alpha.
\end{align*}
Now we know that the rank of $\Sigma_\text{DHC}$ is the same as the rank of $R_\text{DHC}$ and that $\alpha$ and $\beta$ must be full rank because they are changes of the basis matrices, so we have
\begin{align*}
    \text{rank}(R_\text{DHC})=\text{rank}\left(\Sigma-2U_b^TR_{b1}V_p\right).
\end{align*}

This result shows us that the rank of $R_\text{DHC}$ comes from perturbing the rectangular matrix with a typically dense matrix. This means that given an amorphous packing, it is extremely likely for the rank of $R_\text{DHC}$ to be $Nd.$ If the original system is jammed and $\text{rank}(R_\text{DHC})=Nd,$ then 
\begin{align*}
\text{rank}(R_\text{D})=\text{rank}(R)+\text{rank}(R_\text{DHC})=Nd-d.
\end{align*}
From the rank-nullity theorem, we know that the duplicated system only has $d$ floppy modes. This argument for duplicated systems can be applied repeatedly to show that typical unstressed systems with $S\geq d$ states of self-stress will remain rigid when tiled.

\subsection*{Discussion}
\begin{figure}[]
\includegraphics[width=0.475\textwidth]{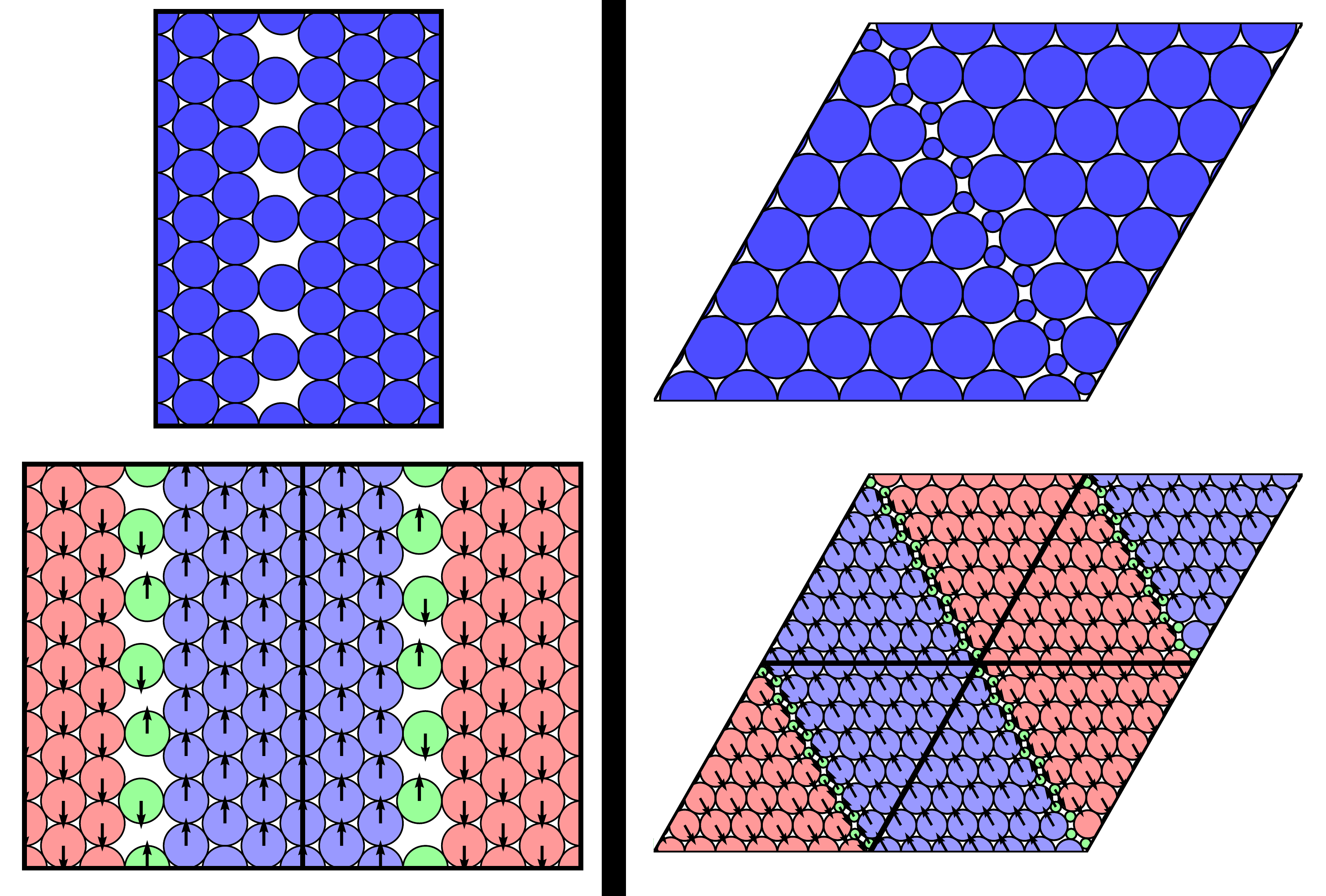}
\caption{\textbf{Top Left}: A jammed packing of hard spheres with more than $d$ states of self-stress based on a triangular lattice with a vertical line of particles that only have three contacts each.\newline
\textbf{Bottom Left}: The corresponding duplicated packing which is not jammed. There is a floppy mode in which the red and blue particles move in opposite directions.\newline
\textbf{Top Right}: Another jammed packing of hard spheres with significantly more than $d$ states of self-stress. This packing is jammed when duplicated once in either direction.\newline
\textbf{Bottom Right}: The packing from above but replicated in a $2 \times 2$ arrangement. This packing is not jammed as the red and blue regions are free to move in opposition, creating a new floppy mode.}\label{fig:breakIt}
\end{figure}

While we have explained why most amorphous, unstressed systems with $d$ or more states of self-stress are typically jammed upon replication, it is worth noting that this does indeed hinge on a statistical argument. It is possible to create non-amorphous packings of hard spheres which are not jammed upon replication. In Fig.~\ref{fig:breakIt}, we create two packings based on the triangular lattice. These packings were proven to be jammed by using a linear programming algorithm~\cite{donev_jamming_2004, donev_linear_2004, torquato_toward_2007}. However, when these packings are tiled, one finds that novel floppy modes are introduced.

\section*{Theorem III:}
\textbf{
The elastic moduli for a jammed packing are the same as the corresponding packing that is duplicated across boundary $x$ up to a trivial scaling factor. This means that the stiffness matrix is extensive when tiling the space.
}
\subsection*{Definitions}
The elastic properties of a packing can be understood from the stiffness matrix, $C,$ where
\begin{align}
\vec{\sigma}=C\vec{\eps}\label{eq:stress-strain}
\end{align}
for the stress, $\vec{\sigma},$ and the strain, $\vec{\eps}.$
We can find this relationship for the original packing by considering the extended Hessian~\cite{dennis_dionysian_2022}.  We define the extended hessian as being the second derivative of the energy function with respect to not only the positional degrees of freedom but the strain degrees of freedom as well. Let $H$ be the extended Hessian such that
\begin{align}
H\equiv \begin{pmatrix}
H_{xx} & H_{x\eps} \\
H_{x\eps}^T & H_{\eps\eps}
\end{pmatrix}
\end{align}
where $H_{xx}$ is the second derivatives of the energy with respect to the positions, $H_{\eps\eps}$ is the second derivatives with respect to strain, and $H_{x\eps}$ is the mixed second derivatives. One can then perform a Taylor expansion of the energy function to arrive at Hooke's law for the extended Hessian,
\begin{align}
H\begin{pmatrix}
\vec{x} \\ \vec{\eps}
\end{pmatrix}=\begin{pmatrix}
-\vec{F} \\ \vec{\sigma}
\end{pmatrix}
\end{align}
where $\vec{F}$ represents the interparticle forces. If we want to find the stress-strain relationship as in Eq.~\ref{eq:stress-strain}, we need to ensure that through the process of applying a strain, the force balance is never lost. This is equivalent to minimizing the energy of a packing after each affine strain step. Therefore, when applying a strain, we also need to apply a non-affine perturbation, $\vec{x}_{\textrm{na}},$ so that
\begin{align}
H\begin{pmatrix}
\vec{x}_{\textrm{na}} \\ \vec{\eps}
\end{pmatrix}&=\begin{pmatrix}
\vec{0} \\ \vec{\sigma}
\end{pmatrix}.
\end{align}
Equivalently, we need to solve the following system of equations:
\begin{align}
\begin{pmatrix}
H_{xx}\vec{x}_{\textrm{na}}+H_{x\eps}\vec{\eps} \\
H_{x\eps}^T\vec{x}_{\textrm{na}}+H_{\eps\eps}\vec{\eps}\end{pmatrix}&=\begin{pmatrix}
\vec{0} \\ \vec{\sigma}
\end{pmatrix}.
\end{align}
If we solve the first system of equations for $\vec{x}_{\textrm{na}}$ and substitute the solution into the second system of equations, we find that
\begin{align}
C=H_{\eps\eps}-H_{x\eps}^T\left(H_{xx}\right)^{-1}H_{x\eps}
\end{align}
where $\left(H_{xx}\right)^{-1}$ is the Moore-Penrose pseudoinverse~\cite{ben-israel_existence_2003} for the singular matrix $H_{xx}.$

\subsection*{Proof}
Now that we have an expression for the stiffness matrix of the original packing, we need to find the stiffness matrix for the duplicated packing, $C_D.$ If we let a system and its duplicate be $A$ and $B$ respectively, we can express the positional second derivatives of the duplicated system, $H_{Dxx},$ as
\begin{align}
H_{Dxx}=\begin{pmatrix}
H_A & H_B \\
H_B & H_A
\end{pmatrix}
\end{align}
since the order in which we take the partial derivatives with respect to the positions of the systems, $A$ and $B$ is inconsequential due to commutativity. The extended Hessian for the duplicated system is therefore
\begin{align}
H_{D}=\begin{pmatrix}
H_A & H_B & H_{x\eps} \\
H_B & H_A & H_{x\eps} \\
H_{x\eps}^T & H_{x\eps}^T & 2H_{\eps\eps}
\end{pmatrix}.
\end{align}
If we let $\vec{x}_1$ be the non-affine motion of the original system and $\vec{x}_2$ be the non-affine motion of the duplicated system, then
\begin{align}
\begin{pmatrix}
H_A\vec{x}_1+H_B\vec{x}_2+H_{x\eps}\vec{\eps}\\
H_B\vec{x}_1+H_A\vec{x}_2+H_{x\eps}\vec{\eps}\\
H_{x\eps}^T\left(\vec{x}_1+\vec{x}_2
\right)+2H_{\eps\eps}\vec{\eps}
\end{pmatrix}=\begin{pmatrix}
\vec{0} \\ \vec{0} \\ \vec{\sigma}
\end{pmatrix}.
\end{align}
Adding the first two equations gives
\begin{align}
\left(H_A+H_B\right)\left(\vec{x}_1+\vec{x}_2\right)+2H_{x\eps}\vec{\eps}=\vec{0}.
\end{align}
We can solve for $\vec{x}_1 + \vec{x}_2$ by using the fact that $H_{xx}=H_A+H_B.$ We see that
\begin{align}
\left(\vec{x}_1 + \vec{x}_2\right)=-2H_{xx}^{-1}H_{x\varepsilon}\vec{\varepsilon}.
\end{align} Making this substitution into the third equation reveals that
\begin{align}
\vec{\sigma}=2\left[H_{\eps\eps}-H_{x\eps}^{T}\left(H_{xx}\right)^{-1}H_{x\eps}\right]\vec{\eps}
\end{align}
or
\begin{align}
C_D=2C.
\end{align}
We can repeat this argument indefinitely, which means that the stiffness matrix is extensive when we tile space with a jammed packing.

\section*{Conclusions}
Periodic boundary conditions are extensively used in the computational modeling of physical systems as they reduce the impact of finite size effects. However, one needs to be cognizant of the limitations and pitfalls of the implementation of these boundary conditions. We have demonstrated that in unstressed jammed packings of soft athermal particles (or spring networks) when periodic boundary conditions are implemented as a tiling, the resulting tiling does not remain jammed or rigid if the original system has fewer than $d$ states of self-stress. We further develop this idea to show that when there are $S<d$ states of self-stress in a jammed packing (or spring network), duplicating the system across any boundary will introduce at least $d-S$ new zero modes. While these proofs are only valid in the absence of prestresses, the presence of prestresses can make the effect even more dire. When there are prestresses, we show that, in general, amorphous jammed packings of soft spheres typically have unjamming motions even when $S\geq d.$ We then argue that it is the over-constrained, unstressed systems (such as spring networks with zero prestress or jammed packings of hard spheres) that are interesting as they typically have an infinite lattice representation that remains rigid when $S\geq d.$ Although there are atypical counter-examples that we present in this paper. 
We conclude the manuscript with a proof that the bulk elastic properties of an infinitely repeated packing are fully captured by periodic boundary conditions. Through these proofs and arguments, this work comprehensively outlines the advantages and pitfalls of utilizing periodic boundary conditions when studying rigidity.

\section*{Acknowledgements}
 We thank Jayson Paulose for his perspective and useful discussions. This work is supported by the Simons Collaboration on Cracking the Glass Problem via awards 454939 (EIC and RCD) and 348126 (VFH).

\bibliography{repetitiveAmoStructures}

\end{document}